\documentclass[10pt,conference,letterpaper]{IEEEtran}
\usepackage{color,dotlessi,graphicx,amsmath,cite,hyperref,amssymb}
\usepackage[spaces,hyphens]{xurl}
\usepackage{booktabs}

\newtheorem{definition}{Definition}
\newtheorem{proposition}{Proposition}
\newtheorem{lemma}{Lemma}
\newtheorem{proof}{Proof}

\newcommand{\mat}[1]{{\boldsymbol{{#1}}}} 

\begin{document}
\title{Response time in a pair of processor sharing queues with Join-the-Shortest-Queue scheduling}
\author{\IEEEauthorblockN{Julianna Bor}
\IEEEauthorblockA{Department of Computing\\
Imperial College London\\
London, UK\\
Email: jb1316@ic.ac.uk}
\and
\IEEEauthorblockN{Peter G Harrison}
\IEEEauthorblockA{Department of Computing\\
Imperial College London\\
London, UK\\
Email: pgh@ic.ac.uk}
}

\maketitle

\begin{abstract}
Join-the-Shortest-Queue (JSQ) is the scheduling policy of choice for many network providers, cloud servers and traffic management systems, where individual queues are served under processor sharing (PS) queueing discipline.
A numerical solution for the response time distribution in two parallel PS queues with JSQ scheduling is derived for the first time.
Using the generating function method, two partial differential equations (PDEs) are obtained corresponding to conditional response times, where the conditioning is on a particular traced task joining the first or the second queue. These PDEs are functional equations that contain partial generating functions and their partial derivatives, and therefore cannot be solved by commonly used techniques.
We are able to solve these PDEs numerically with good accuracy and perform the deconditioning with respect to the queue-length probabilities by evaluating a certain complex integral.
Numerical results for the density and the first four moments compare well against regenerative simulation with 500,000 regeneration cycles.
\end{abstract}

\begin{IEEEkeywords}
Queueing theory, response times, generating functions.
\end{IEEEkeywords}

\section{Introduction}~\\
Join-the-Shortest-Queue (JSQ) is a scheduling policy that assigns arriving jobs to the shortest of a number of parallel queues and breaks ties by assigning jobs randomly with a given probability. JSQ---also called Least-Connection scheduling in a load balancing context---is the policy of choice for many network providers, cloud servers and traffic management systems, where individual queues are served under processor sharing (PS) queueing discipline. 
For example, JSQ is offered by Cisco IOS Server~\cite{cisco}, Azure Applicaton Gateway~\cite{azure}, Kubernetes~\cite{kubernetes} and AWS Application Load Balancer~\cite{aws}.
Despite the ubiquity of JSQ scheduling, there is still a lot unknown regarding its behaviour quantitatively. 

Consider a system of two parallel queues where Poisson arrivals occur at rate $\lambda$ and join the shorter of the two queues, or, if the lengths are equal, join queue $1$ with probability $a_1$ and queue $2$ with probability $a_2=1-a_1$. Each queue has an independent server with exponential service times, parameters $\mu_1, \mu_2$. In other words, consider a pair of parallel $M/M/1$ queues with JSQ scheduling. This apparently simple problem was first considered by Kendall almost 70 years ago in the symmetric case when the queueing discipline is FCFS \cite{kin1961}, i.e. where $a_1=a_2=1/2$ and $\mu_1=\mu_2$. 
Subsequently, Flatto and McKean obtained a complex closed form solution for the queue-length distribution and mean response time of this case in \cite{FlaMcK1977}.

In a FCFS system, once a customer joins the queue, no later arriving customer can overtake it. This means that in order to calculate the cumulative distribution function of the time it takes the customer to leave the queue, there is no need to track the customers joining the queue after said customer.
Therefore the calculation of response time distribution in a FCFS system is straightforwardly reduced to obtaining just the equilibrium queue length probabilities.
Functional equations leading to the generating function of the equilibrium queue length probabilities for a pair of JSQ-FCFS queues have been obtained and solved recently in~\cite{pgh2023}.

Now, take Kendall's model and modify it ever so slightly by replacing the FCFS queueing discipline with PS at the individual queues. 
This is a natural next step, as PS is often the preferred queueing discipline in practice. 
PS and FCFS have the same mean response time but the higher moments of PS are larger so that users experience greater variability.
However, PS favours short jobs considerably and a user's mean response time at equilibrium is proportional to its own service requirement~\cite{Asare1983}.
This inherent fairness is the reason why PS tends to be favoured over FCFS.
Obtaining the response time distribution of the PS version of Kendall's `simple' problem above remained unsolved for decades and is the focus of this paper. 
In fact, we solve a more general version of Kendall's model with PS queueing discipline.

The rest of the paper is organized as follows. Functional equations for the Laplace-Stieltjes transform (LST) of the response time distribution are obtained in Section \ref{jsq}. The functional equations are transformed into a problem in linear algebra and subsequently solved in Section \ref{funceq-sol}.
The method is evaluated in Section \ref{evaluation}. Conclusions are drawn in Section \ref{conc}.

\section{Functional equations for the response time of two JSQ-PS queues}
\begin{figure}[ht]
    \centering
    \includegraphics[scale=0.35]{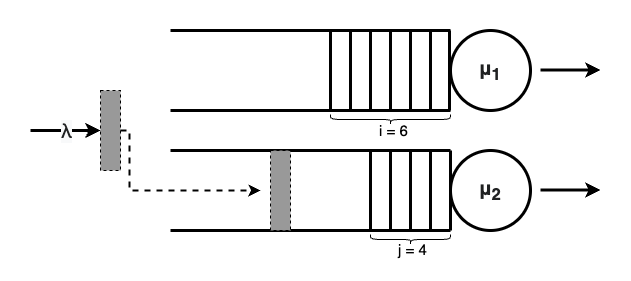}
    \caption{Two JSQ-PS queues in parallel where the arriving task joins the shorter queue. The arrival rate is $\lambda$, the service rates are $\mu_1$ and $\mu_2$, }
    \label{jsq-network}
\end{figure}

~\\
\label{jsq}
As outlined above, the queueing model under consideration consists of two parallel queues each with exponential service times and service rates of $\mu_1$ and $\mu_2$, respectively. Both queues use PS discipline locally. Arrivals are Poisson with rate $\lambda$, assuming $\lambda < \mu_1+\mu_2$ for stability, and join the shortest queue. If at an arrival instant the queues are of equal length, the arriving job is routed to queue 1 with probability $a_1$ and queue 2 with probability $a_2=1-a_1$.
Figure \ref{jsq-network} displays the above network at an arrival instant, when queue 1 has length $i=6$ and queue 2 has length $j=4$. The arriving task is therefore joining queue 2.

The goal is to derive the LST of the response time probability distribution, which is then inverted numerically to obtain the distribution (or density).
In order to do that, we first derive two functional equations. 
These functional equations are then solved in Section \ref{funceq-sol}.

\subsection{Generating function for the queue-length distribution}
To deal with response times at equilibrium, one first needs to obtain the joint steady-state queue-length distribution. This is derived for JSQ servers with FCFS queueing discipline in \cite{pgh2023}, but PS has exactly the same balance equations for the queue length probabilities in a Markovian model, and hence the same solution. 
Let $G(x,y)$ denote the generating function of the steady state queue-length probabilities, that is $G(x,y)=\sum\limits_{i=0}^\infty\sum\limits_{j=0}^\infty \pi_{i,j}x^iy^j$ where $\pi_{i,j}$ is the probability of $i$ and $j$ customers in the first and second queue, respectively, at equilibrium. In this paper we assume $G(x,y)$ is known.

\subsection{Generating functions of conditional response time densities}
Our next step is to obtain functional equations for the generating functions of the LST of response time distributions, conditioned on whether the customer is in the first or the second queue.

\noindent Let the random variables $U_{i,j}$, for $i,j\geq 0$, denote the remaining time to completion of service of a particular customer in queue 1 when there are $i$ and $j$ other customers present in queues 1 and 2, respectively.  Let $U_{i,j}$ have probability distribution function $U_{i,j}(t)$, with LST $U_{i,j}^*(s)$.  We define $V_{i,j},V_{i,j}(t)$ and $V_{i,j}^*(s)$ analogously for a tagged customer in queue 2.  
We denote the generating function of $U_{i,j}^*(s)$ by $E(x,y,s)$ and the generating function of $V_{i,j}^*(s)$ by $F(x,y,s)$.

In order to derive functional equations for $E(x,y,s)$ and $F(x,y,s)$, we define certain related Taylor series, each holomorphic in unit disks centered at the origin.  Some of them are termed ``partial generating functions" because they are power series in two variables that are not summed over the full first quadrant $\{(i,j) \mid 0\leq i < \infty, 0 \leq j < \infty\}$.
We define these related Taylor series in the following definition.
\smallskip
\begin{definition}
\label{functionDefs}
Generating functions corresponding to $U_{i,j}^*(s)$ and $V_{i,j}^*(s)$ and related analytic functions:
\begin{enumerate}
\item Generating functions

$$E(x,y,s):=\sum\limits_{i=0}^\infty \sum\limits_{j=0}^\infty U_{i,j}^*(s)x^iy^j$$
and
$F(x,y,s)$ is defined similarly for $V_{i,j}^*(s)$.
\item Partial generating functions

$$E^\leq(x,y,s):=\sum\limits_{i=0}^\infty \sum\limits_{j=i}^\infty U_{i,j}^*(s)x^iy^j$$

Analogous expressions are defined for $E^\geq$ and for $F^\leq$, $F^\geq$, replacing $U$ by $V$ on the right hand side.
\item Related analytic functions
\vspace{-3mm}
\begin{align*}
    \alpha_E(x,y,s)&:=\sum_{i=0}^\infty \sum_{j=0}^\infty U^*_{i+1,j}(s) x^iy^j\\ 
    \beta_E(x,y,s)&:=\sum_{i=0}^\infty \sum_{j=0}^\infty U^*_{i,j+1}(s) x^iy^j\\
    \gamma_E(x,y,s)&:=\sum_{i=0}^\infty \sum_{j=0}^\infty U^*_{i+2,j}(s) x^iy^j\\ 
    \delta_E(x,y,s)&:=\sum_{i=0}^\infty \sum_{j=0}^\infty U^*_{i,j+2}(s) x^iy^j
\end{align*}
and $\alpha_F(x,y,s),\beta_F(x,y,s),\gamma_F(x,y,s),\delta_F(x,y,s)$ are defined by replacing $U$ by $V$ on the right hand sides.
\item Partial versions of related analytic functions
\vspace{-3mm}
\begin{align*}
    \alpha^{\leq}_E(x,y,s) &= \sum_{i=0}^\infty \sum_{j=i}^\infty U^*_{i+1,j}(s) x^iy^j\\
    \alpha^{\geq}_E(x,y,s) &= \sum_{i=0}^\infty \sum_{j=0}^{i} U^*_{i+1,j}(s) x^iy^j
\end{align*}
Similarly for $\beta^{\leq}_E,\beta^{\geq}_E,\gamma^{\leq}_E,\gamma^{\geq}_E, \delta^{\leq}_E, \delta^{\geq}_E$ and
for $\alpha_F,\beta_F,\gamma_F,\delta_F$ by replacing $U$ by $V$.
\end{enumerate}
\end{definition}
\smallskip
Generalizing the notation for partial generating functions, for $f(x,y) = \sum_{i=0}^\infty \sum_{j=0}^\infty a_{i,j}x^iy^j$ and for binary operator $\sim$, we write 
$$f^\sim  = \sum_{i \sim j} a_{i,j}x^iy^j$$
Note that $f^\sim$, where $\sim$ stands for one of $<,>,=$, can be expressed using $f^\leq(x,y)$,$f^\geq(x,y)$ and $f(x,y)$. For example, $f^=(x,y) = f(x,y)^\leq + f(x,y)^\geq - f(x,y)$. 

The following lemma shows how the related analytic functions can be expressed in terms of the original generating functions $E$ and $F$.
It is easy to derive by appropriate change of variable over the summation domains.

\begin{lemma}
{\small
\begin{align*}
        \alpha_E(x,y,s)&=\frac 1 x\left[E(x,y,s)-E(0,y,s)\right]\\ \gamma_E(x,y,s)&=\frac 1 {x^2}\left[E(x,y,s)-E(0,y,s)-x \frac{\partial E}{\partial x}(0,y,s)\right]\\
        \beta_E(x,y,s)&=\frac 1 y\left[E(x,y,s)-E(x,0,s)\right]\\
        \delta_E(x,y,s)&=\frac 1 {y^2}\left[E(x,y,s)-E(x,0,s)-y \frac{\partial E}{\partial y}(x,0,s)\right]
    \end{align*}
}%
Analogous expressions can be obtained for $\alpha_F,\beta_F,\gamma_F,\delta_F$,
replacing $E$ by $F$ on the right hand sides.
\label{relatedAnalFunctions}
\end{lemma}

Partial generating functions can be computed as follows~\cite{pgh2023}.

$$f^\leq (x_0,y_0) = \frac{1}{2\pi i} \oint_{C_y} \frac{f(x_0y_0/y,y)}{y-y_0} dy \label{gleq}$$
Again, similar expression can be derived for $f^\geq$.

We now have the following result for the generating functions $E$ and $F$ corresponding to the LST of conditional response time distributions for a task joining queue 1 and queue 2, respectively.
\medskip
\begin{proposition}
\label{prop:EandF}
The generating functions $E(x,y,s)$ and $F(x,y,s)$ are given by the following equations:
{\small
\begin{equation}
\begin{split}
\big(s+&\lambda+(1-x)\mu_1+(1-y)\mu_2\big) \big(E(x,y,s)+x \frac{\partial E}{\partial x}(x,y,s)\big)\\
& - \mu_2 \big(E(x,0,s)+x \frac{\partial E}{\partial x}(x,0,s)\big) - \lambda \frac{\partial E^{<}}{\partial x}(x,y,s)\\ 
& - \lambda \big(\beta_E^\geq(x,y,s) + x \frac{\partial \beta_E^\geq}{\partial x}(x,y,s)\big) - \lambda a_1 \frac{\partial E^=}{\partial x}(x,y,s)\\ 
& - \lambda a_2 y \big(\delta_E^=(x,y,s) + x \frac{\partial \delta_E^=}{\partial x}(x,y,s)\big)
= \frac{\mu_1}{(1-x)(1-y)}
\end{split}
\label{eqforE}
\end{equation}
}%
and
{\small
\begin{equation}
\begin{split}
\big(s+&\lambda+(1-x)\mu_1+(1-y)\mu_2\big) \big(F(x,y,s)+y \frac{\partial F}{\partial y}(x,y,s)\big)\\
& - \mu_1 \big(F(0,y,s)+y \frac{\partial F}{\partial y}(0,y,s)\big) - \lambda \frac{\partial F^>}{\partial y}(x,y,s) \\
&- \lambda \big(\alpha_F^\leq(x,y,s) + y \frac{\partial \alpha_F^\leq}{\partial y}(x,y,s)\big) - \lambda a_2 \frac{\partial F^=}{\partial y}(x,y,s)\\
& - \lambda a_1 x \big(\gamma_F^=(x,y,s) + y \frac{\partial \gamma_F^=}{\partial y}(x,y,s)\big)
= \frac{\mu_2}{(1-x)(1-y)}
\end{split}
\label{eqforF}
\end{equation}
}%
\end{proposition}
The proof is omitted here due to space constraints
.
So far, we have derived functional equations for $E$ in Equation \eqref{eqforE} and $F$ in Equation \eqref{eqforF}. Our next task is to combine these generating functions in order to express the LST of unconditional response time distribution.

\subsection{LST of the unconditional response time distribution}
The functional equations \eqref{eqforE} and \eqref{eqforF} are solved for generating functions $E$ and $F$ in section \ref{funceq-sol}.
However, it is still not straightforward to express the LST of unconditional response time distributions in terms of the above generating functions. 
This is because in the case of PS discipline overtaking might occur. Overtaking means that a customer that joins a queue can finish its service before some or all the customers that were already in the queue at the time of its arrival.
It has a crucial effect as the rate at which a customer is being served at either queue at any one time is dependent on the number of customers in its queue---given $i$ customers in queue 1, say, the rate at which a single customer is served is $\mu_1/i$.  The value of $i$ in turn is dependent on the length of the other queue since arriving jobs always join the shorter of the two queues. 
Calculation of response time distribution under PS is therefore significantly more complicated than for FCFS. PS disclipline requires the tracking of later arrivals to either queue whereas FCFS does not.
As a result, the conditional LSTs are not geometric expressions and so more work is needed to perform the deconditioning.  The required Laplace transform is given by the following proposition.

\medskip
\begin{proposition}
\label{wstarprop}
The Laplace transform of response time density in a pair of M/M/1 queues with JSQ-PS scheduling, omitting $s$ from the integrands for brevity, is 
{\small
\begin{align}
    &W^*(s)= \frac{1}{(2\pi)^2} \nonumber\\ 
      &\Bigg( \int_0^{2\pi}\int_0^{2\pi} E^<(r_1e^{\dotlessi t_1},r_1e^{\dotlessi t_2})G(r_2e^{-\dotlessi t_1},r_2e^{-\dotlessi t_2}) dt_1 dt_2 + \nonumber \\
    &\int_0^{2\pi}\int_0^{2\pi}F^>(r_1e^{\dotlessi t_1},r_1e^{\dotlessi t_2}) G(r_2e^{-\dotlessi t_1},r_2e^{-\dotlessi t_2}) dt_1 dt_2 + \nonumber \\
    & \int_0^{2\pi} \int_0^{2\pi}a_1E^=(r_1e^{\dotlessi t_1},r_1e^{\dotlessi t_2}) G(r_2e^{-\dotlessi t_1},r_2e^{-\dotlessi t_2}) dt_1 dt_2+ \nonumber \\
    &\int_0^{2\pi} \int_0^{2\pi} a_2F^=(r_1e^{\dotlessi t_1},r_1e^{\dotlessi t_2})G(r_2e^{-\dotlessi t_1},r_2e^{-\dotlessi t_2}) dt_1 dt_2 \Bigg)
    \label{wstareq}
\end{align}
}%
where $r_1 \leq 1, r_2 = 1/r_1 \leq \frac{\mu_1+\mu_2}{\lambda}$ and $\dotlessi$ denotes the imaginary unit.
\end{proposition}

In this procedure, the choices of $r_1$ and $r_2$ are important. The conditional Laplace transforms' generating functions ($E(x,y,s)$ and $F(x,y,s)$) diverge at $r_1=1$ when $s=0$, each then being equal to $1/((1-x)(1-y))$.  Similarly, $G(x,y)$ diverges at its radius of convergence, which we took to be $(\mu_1+\mu_2)/\lambda$.  We chose a value for $r_2$ halfway between 1 and $G$'s radius of convergence, i.e. $r_2=\frac{\lambda+\mu_1+\mu_2}{2\lambda}$.  However, note that for heavily loaded systems, $\lambda \simeq \mu_1+\mu_2$ so that $r_1$ ends up near to 1 and $r_2$ near to $G$'s radius of convergence.  Hence, significant numerical errors can be expected at high load.


\section{Solving the functional equations}\label{funceq-sol}
In Section \ref{jsq}, we stated three functional equations for the generating functions of conditional and unconditional response time distribution LSTs, namely equations \eqref{eqforE}, \eqref{eqforF} and \eqref{wstareq}. 
To solve these functional equations numerically. we transform them into a problem in linear algebra.

\begin{figure}[ht]
    \centering
    \includegraphics[scale=0.5]{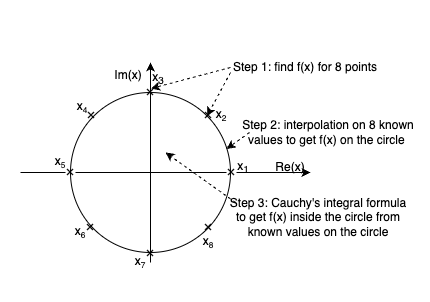}
    \caption{Steps to approximate f(x) from 8 known values around the unit circle.}
    \label{linalgproblem}
\end{figure}

Consider a bivariate functional equation of the form:
{\small
\begin{equation} \label{exemplar-fn-eq}
a(x,y)G(x,y)=b(x,y)+c(x,y)G(x,0)+d(x,y)\Delta_yG(0,y)
\end{equation}
}%

 where $a(x,y)$, $b(x,y)$, $c(x,y)$ and $d(x,y)$ are known functions of $x$ and $y$, and $\Delta_y$ denotes partial differentiation wrt $y$. This is a complex equation that might be solvable by Boundary Value Problem (BVP) methods, but a numerical solution would appear problematic. Our numerical approximation method is the following.

Let $C_x$ and $C_y$ be contours in the complex planes of $x$ and $y$. Typically, $C_x$ and $C_y$ are circles. The interior of $C_x$ is then a disk $D_x$, and similarly for $C_y, D_y$.
Now, the value of a function $f(x,y)$ at any point $(x_0,y_0)$ inside the Cartesian product of the contours, $C_x \times C_y$ (i.e. with $x_0 \in D_x$, and $y_0 \in D_y$), can be obtained by Cauchy's Integral Theorem applied twice in two dimensions, provided $f(x,y)$ is known for all $x \in C_x, y \in C_y$; see \cite{cauchySeveralVarPolydisc,priestley} for example.  
We therefore first seek an interpolation for $f$ over $\{(x,y) \in C_x \times C_y\}$ by choosing two sets of discretization points $x_1,\ldots,x_m \in C_x$ and $y_1,\ldots,y_n \in C_y$.  We take $m=n$ to simplify the notation; the analysis is no more complicated when $m \neq n$ but is more cluttered.
Choosing $C_x$ and $C_y$ to be circles means the interpolation can be done easily on the real-valued arguments (angles) of the discretization points, giving more precision. Moreover, interpolation on reals is directly supported in most mathematical software packages. 
An approximation to $f(x,b)$ on $C_x$, for fixed $b \in C_y$, is an interpolation over the points $x_1,\ldots,x_n$ of the values $f(x_1,b),\ldots,f(x_n,b)$, and similarly for $f(a,y)$ at fixed $a\in C_x$.  
We order the point-pairs alphabetically by subscript, $\{(x_1,y_1),\ldots,(x_1,y_n),(x_2,y_1),\ldots,(x_n,y_n)\}$ and say that $\vec f = (f((x_1,y_1),\ldots,f(x_1,y_n),f(x_2,y_1),\ldots,f(x_n,y_n))$ is a \emph{vector representation} of $f$ on $C_x \times C_y$. The vector representation can even be symbolic, i.e. $\vec f = (f_{11},\ldots,f_{nn})$.  
Figure \ref{linalgproblem} shows how the approximation works in one dimension.
If now we can find matrices to represent the operators in the functional equation being solved -- called an operator's \emph{matrix representation} -- we can rewrite the functional equation as a matrix-vector equation.
\begin{table}[t]
\caption{Matrices representing operations in the summands of the functional equation for $G(x,y)$.}
\begin{center}
\begin{tabular}{cc}
\toprule
Operation & Matrix \\
\midrule
Identity operator giving $G$ & $\mat I^{nm \times nm}$ \\
Multiplication by known function $A(x,y)$ & $\mat \Lambda(A)$\\
Differentiation wrt $x$ & $\mat M_{1,0}$ \\
Differentiation wrt $y$ & $\mat M_{0,1}$ \\
Values on $C_y$ at a point $u_0 \in D_x$ & $\mat U_{u_0}$ \\
Values on $C_x$ at a point $v_0 \in D_y$ & $\mat V_{v_0}$ \\
Partial generating function $G^\leq $ &  $\mat P^\leq $ \\
Partial generating function $G^\geq $ &  $\mat P^\geq $ \\
\bottomrule
\end{tabular}
\end{center}
\label{matrix-map}
\end{table}
For example, multiplication by a fixed function $a(x,y)$ is represented by a diagonal matrix of the function's values at the discretization point-pairs, ordered alphabetically,
 and we define
 $$ \mat \Lambda(\vec a) = \textrm{Diag}\Big(a(x_i,y_j)\mid 1\leq i \leq n, 1 \leq j \leq n\Big).$$  
Matrices for the operators of differentiation and partialization (i.e. producing a partial generating function) were obtained in \cite{pgh2023} by considering each point $x_1,\ldots,x_n, y_1,\ldots,y_n$ in turn, distorting the contour to go around it, and applying Cauchy's Integral Theorem.  The details are not reproduced here, apart from Table \ref{matrix-map} which lists the matrices corresponding to each operator of interest.  Notice that the matrices for differentiation and partialization are constant, applying to all functional equations with $n$ discretization points in both complex planes.
Equation \eqref{exemplar-fn-eq} above is thereby transformed to:
$$ (\mat\Lambda(\vec a)-\mat\Lambda(\vec c)\mat V_0-\mat\Lambda(\vec d)\mat M_{0,1}\mat U_0)\vec g=\mat\Lambda(\vec b)
$$
giving a vector representation $$\big(\mat\Lambda(\vec a)-\mat\Lambda(\vec c)\mat V_0-\mat\Lambda(\vec d)\mat M_{0,1}\mat U_0\big)^{-1} \mat \Lambda(\vec b)$$ for the solution-function $G(x,y)$ on the contours.

\subsection*{Conditional response times}
In order to transform the equations for functions $E(x,y,s)$ and $F(x,y,s)$, we need to define new matrices for the related analytic functions of Definition \ref{functionDefs}.
\smallskip
\begin{proposition}
\label{prop:additional-mx}
    Let $\alpha(x,y)$ denote the operation performed on function $f$ to get $\alpha_f(x,y)$ and let $\mat A$ be the matrix representation of operation $\alpha(x,y)$. Define $\mat B, \mat C$ and $\mat D$ in a similar manner for the operations $\beta(x,y), \delta(x,y)$ and $\gamma(x,y)$, respectively. The matrices $\mat A,\mat B, \mat C$ and $\mat D$ are given by:
    {\small
    \begin{align*}
        &\mat A = \mat\Lambda(1/\vec x)\big(\mat I^{n^2\times n^2}-\mat U_0\big)\\
        & \mat B = \mat\Lambda(1/\vec y)\big(\mat I^{n^2\times n^2}-\mat V_0\big)\\
        &\mat C = \mat\Lambda(1/\vec x^2)
    (\mat I^{n^2\times n^2}-\mat U_0-\mat\Lambda(\vec x)\mat U_0\mat M_{1,0}) \\
        &\mat D = \mat\Lambda(1/\vec y^2)
    (\mat I^{n^2\times n^2}-\mat V_0-\mat\Lambda(\vec y)\mat V_0\mat M_{0,1})
    \end{align*}
    }%
\end{proposition}

\begin{table}[t]
\caption{Matrices representing the operations corresponding to the related analytic functions of Definition \ref{functionDefs}.}
\begin{center}
\begin{tabular}{cc}
\toprule
Operation & Matrix \\
\midrule
$\alpha(x,y) = \frac 1 x\left[f(x,y)-f(0,y)\right]$ & $\mat A^{nm \times nm}$ \\
$\beta(x,y) = \frac 1 y\left[f(x,y)-f(x,0)\right]$ & $\mat B^{nm \times nm}$ \\
$\gamma(x,y) = \frac 1 {x^2}\left[f(x,y)-f(0,y)-x \frac{\partial}{\partial x}f(0,y)\right]$ & $\mat C^{nm \times nm}$ \\
$\delta(x,y) = \frac 1 {y^2}\left[f(x,y)-f(x,0)-y \frac{\partial}{\partial y}f(x,0)\right]$ & $\mat D^{nm \times nm}$ \\
\bottomrule
\end{tabular}
\end{center}
\label{matrix-map-additional}
\end{table}

Equipped with these new matrices, the transformation of the equations for $E(x,y,s)$ and $F(x,y,s)$, albeit more complicated, follows the same procedure introduced above.
For the vector representation $\vec e$ of $E(x,y,s)$, by replacing operators by their corresponding matrices in Table \ref{matrix-map} and Table \ref{matrix-map-additional}, we obtain from Equation \eqref{eqforE}:
{\small
\begin{align} \label{E-vec-rep-eq}
\Big(\mat\Lambda(s&+\lambda+(1-\vec x)\mu_1+(1-\vec y)\mu_2)\big(\mat I^{n^2\times n^2}
    +\mat\Lambda(\vec x)\mat M_{1,0} \big)\nonumber \\ 
    &- \mu_2 \big(\mat V_0+\mat\Lambda(\vec x)\mat V_0\mat M_{1,0}\big)- \lambda \mat M_{1,0}(\mat I^{n^2\times n^2}-\mat P^\geq)\nonumber\\
    &-\lambda\big(\mat I^{n^2\times n^2}+\mat\Lambda(\vec x)
    \mat M_{1,0} \big) \mat P^\geq\mat B - \lambda a_1 \mat M_{1,0}\mat P^=\nonumber\\
    & - \lambda a_2 \mat\Lambda(\vec y)(\mat I^{n^2\times n^2}+\mat\Lambda(\vec x)\mat M_{1,0} \big)\mat P^= \mat D \Big)\vec e = \nonumber\\
    &\left(\frac{\mu_1}{(1-x_i)(1-y_j)} \mid 1\leq i \leq n,1\leq j\leq n\right)
\end{align}
}%
where the vector on the right hand side is ordered alphabetically as usual.
Similarly, for the vector representation $\vec f$ of $F(x,y,s)$, we obtain
{\small
\begin{align} \label{F-vec-rep-eq}
\Big(\mat\Lambda(s&+\lambda+(1-\vec x)\mu_1+(1-\vec y)\mu_2)\big(\mat I^{n^2\times n^2}
    +\mat\Lambda(\vec y)\mat M_{0,1} \big)\nonumber \\ 
    &- \mu_1 \big(\mat U_0+\mat\Lambda(\vec y)\mat U_0 \mat M_{0,1}\big) - \lambda \mat M_{0,1}(\mat I^{n^2\times n^2}-\mat P^\leq)\nonumber\\
    &-\lambda\big(\mat I^{n^2\times n^2}+\mat\Lambda(\vec y)
    \mat M_{0,1} \big)\mat P^\leq\mat A - \lambda a_2 \mat M_{01}P^=\nonumber\\ 
    &- \lambda a_1 \mat\Lambda(\vec x)(\mat I^{n^2\times n^2}+\mat\Lambda(\vec y)\mat M_{0,1} \big) \mat P^= \mat C \Big)\vec f = \nonumber\\
    &\left(\frac{\mu_2}{(1-x_i)(1-y_j)} \mid 1\leq i \leq n,1\leq j\leq n\right) 
\end{align}
}%
We can now approximate the function $E(x,y,s)$ by an interpolation of the vector representation $\mat \Gamma_E \Big(\frac{\mu_1}{(1-x_i)(1-y_j)} \mid 1\leq i,j \leq n\Big)$ of $E(x,y,s)$ over the discretization set pairs $\{(x_i,y_j) \mid 1\leq i,j \leq n\}$,
where
{\small
\begin{align}  \label{gammaEdef}
\mat \Gamma_E &= \Big(\mat\Lambda(s+\lambda+(1-\vec x)\mu_1+(1-\vec y)\mu_2)\big(\mat I^{n^2\times n^2} +\mat\Lambda(\vec x)\mat M_{1,0} \big)\nonumber \\
    &- \mu_2 \big(\mat V_0+\mat\Lambda(\vec x)\mat V_0\mat M_{1,0}\big)- \lambda \mat M_{1,0}(\mat I^{n^2\times n^2}-\mat P^\geq) \nonumber \\
    &-\lambda\big(\mat I^{n^2\times n^2}+\mat\Lambda(\vec x)
    \mat M_{1,0} \big) \mat P^\geq\mat B- \lambda a_1 \mat M_{1,0}\mat P^= \nonumber \\
    &- \lambda a_2 \mat\Lambda(\vec y)(\mat I^{n^2\times n^2}+\mat\Lambda(\vec x)\mat M_{1,0} \big)\mat P^= \mat D \Big)^{-1}
\end{align}
}%
Similarly, $F(x,y,s)$ is approximated by an interpolation of the vector representation $\mat \Gamma_F \Big(\frac{\mu_2}{(1-x_i)(1-y_j)} \mid 1\leq i,j \leq n\Big)$ of $F(x,y,s)$ over the discretization set pairs $\{(x_i,y_j) \mid 1\leq i,j \leq n\}$,
where
{\small
\begin{align} \label{gammaFdef}
\mat \Gamma_F &= \Big(\mat\Lambda(s+\lambda+(1-\vec x)\mu_1+(1-\vec y)\mu_2)\big(\mat I^{n^2\times n^2}+\mat\Lambda(\vec y)\mat M_{0,1} \big)\nonumber \\
    &- \mu_1 \big(\mat U_0+\mat\Lambda(\vec y)\mat U_0 \mat M_{0,1}\big) - \lambda \mat M_{0,1}(\mat I^{n^2\times n^2}-\mat P^\leq) \nonumber \\
    &-\lambda\big(\mat I^{n^2\times n^2}+\mat\Lambda(\vec y)
    \mat M_{0,1} \big)\mat P^\leq\mat A - \lambda a_2 \mat M_{01}P^= \nonumber \\
    &- \lambda a_1 \mat\Lambda(\vec x)(\mat I^{n^2\times n^2}+\mat\Lambda(\vec y)\mat M_{0,1} \big)\mat P^= \mat C \Big)^{-1}
\end{align}
}%
One-variable interpolation is provided in most mathematical software packages, but our two-variable interpolation routine, in Mathematica code, may be of interest.

\subsubsection{Unconditional response time} 
Let $\vec e^< = (\mat I - \mat P^{\geq})\vec e$, $\vec f^>=(\mat I - \mat P^{\leq})\vec f, \vec e^= = \mat P^= \vec e$ and $\vec f^= = \mat P^= \vec f$. Then equation \eqref{wstareq} for $W^*(s)$ in Proposition \ref{wstarprop} can be transformed as follows, omitting $s$ from the integrands for brevity:
{\small
\begin{align*}
    &W^*(s) \approx \frac{1}{(2\pi)^2} \cdot \nonumber \\
    &\Bigg( \int_0^{2\pi} \int_0^{2\pi}\mathfrak I(\vec e^<)(r_1e^{\dotlessi t_1},r_1e^{\dotlessi t_2})\mathfrak I(\vec g)(r_2e^{-\dotlessi t_1},r_2e^{-\dotlessi t_2})dt_1 dt_2 + \nonumber\\
    &\int_0^{2\pi} \int_0^{2\pi} \mathfrak I(\vec f^>)(r_1e^{\dotlessi t_1},r_1e^{\dotlessi t_2})
    \mathfrak I(\vec g)(r_2e^{-\dotlessi t_1},r_2e^{-\dotlessi t_2})dt_1 dt_2 + \\
    & \int_0^{2\pi} \int_0^{2\pi} a_1\mathfrak I(\vec e^=)(r_1e^{\dotlessi t_1},r_1e^{\dotlessi t_2})\mathfrak I(\vec g)(r_2e^{-\dotlessi t_1},r_2e^{-\dotlessi t_2})dt_1 dt_2 + \nonumber \\
    &\int_0^{2\pi} \int_0^{2\pi} a_2\mathfrak I(\vec f^=)(r_1e^{\dotlessi t_1},r_1e^{\dotlessi t_2})
    \mathfrak I(\vec g)(r_2e^{-\dotlessi t_1},r_2e^{-\dotlessi t_2})dt_1 dt_2
    \Bigg)
\end{align*}
}%
where $\mathfrak I(\vec e^\sim)$, $\mathfrak I(\vec f^\sim)$ and $\mathfrak I(\vec g)$ are two-variable interpolations of $\vec e^\sim$, $\vec f^\sim$ and  $\vec g$, respectively, defined above. We used linear interpolation.

One can choose the contours on which the original discretization points lie to match the circles of this double integral that calculates $W^*(s)$ according to equation \eqref{wstareq}. That is, the circles used in the estimation of $E$ and $F$ have radius $r_1$ and the circle used to estimate the pgf $G$ has radius $r_2=1/r_1$.
This leads to a straightforward evaluation of the integral by interpolating the discretized values around the circles.

\subsubsection{Moments}
The $k$th moment of response time is $(-1)^k$ multiplied by the $k$th derivative of $W^*(s)$ evaluated at $s=0$.  This requires the partial derivatives of $E(x,y,s)$ and $F(x,y,s)$ with respect to (wrt) $s$ at $s=0$, which are determined below in Lemma \ref{lem-moms}.

First, note that the solution for the vector representation of $E(x,y,s)$ in equation \eqref{E-vec-rep-eq}, with right hand side changed to $r_E(x,y,s)$, is $\mat \Gamma_E \vec v_E$, where the column vector $\vec v_E = (r_E(x_i,y_j,s)\,|\, 1\leq i,j\leq n)^T$, ordered alphabetically; so in the above, we had $r_E(x_i,y_j,s)=\frac{\mu_1}{(1-x_i)(1-y_j)}$ giving 
$$\vec e = \mat \Gamma_E \left(\frac{\mu_1}{(1-x_i)(1-y_j)}\,|\,1\leq i,j\leq n\right)^T$$
The results for $F(x,y,s)$ are similar.

\begin{lemma} \label{lem-moms} 
The $k$th partial derivative of $E(x,y,s)$ wrt $s$ at $s=0$ has vector representation $\vec e_k$ given by the recurrence:
\begin{align*}
&\vec e_0 = \mat \Gamma_E \left(\frac{\mu_1}{(1-x_i)(1-y_j)}\,|\,1\leq i,j\leq n\right)^T, \\ 
&\vec e_k=-k \mat \Gamma_E \left(\vec I^{n^2\times n^2} + \mat\Lambda(\vec x)M_{1,0}\right) \vec e_{k-1}.
\end{align*}
where $\mat \Gamma_E$ is defined by equation \eqref{gammaEdef} above with the substitution $s=0$.

Similarly, the $k$th partial derivative of $F(x,y,s)$ wrt $s$ at $s=0$ has vector representation $\vec f_k$ given by the recurrence:
\begin{align*}
&\vec f_0 = \mat \Gamma_F \left(\frac{\mu_2}{(1-x_i)(1-y_j)}\,|\,1\leq i,j\leq n\right)^T, \\ 
&\vec f_k=-k \mat \Gamma_F \left(\vec I^{n^2\times n^2} + \mat\Lambda(\vec y)M_{0,1}\right) \vec f_{k-1}.
\end{align*}
where $\mat \Gamma_F$ is defined by equation \eqref{gammaFdef} above with the substitution $s=0$.
\end{lemma}

We note that there is no need to recalculate the inverse of a matrix for each moment; the same inverse matrix $\Gamma_E$ or $\Gamma_F$ is all that is needed.  As a result, once these inverses have been computed, any number of moments can be calculated quickly.

The moments of unconditional response time are obtained by deconditioning according to Proposition \ref{wstarprop}, replacing $W^*(0)$ by the $k$th moment $M_k$, $E$ by $E^{(k)}$ and $F$ by $F^{(k)}$; actually, for $k=0$, $W^*(0)=M_0=1$, providing an accuracy check.

\section{Evaluation}\label{evaluation}
In order to evaluate the methodology presented above for estimating the response time distribution of two JSQ-PS queues, we perform several experiments.
First, the numerical response time moments and density are obtained for two networks, one with medium load and one with high load. The results are then compared against simulation.
Next, JSQ scheduling is compared against static scheduling methods.
This is followed by a section devoted to finding the optimal value of $a_1$, the probability of a customer joining the first queue given that the queues are of equal length.
Finally, the algorithm's accuracy and its computational cost are considered.

\subsection{Comparison with simulation}\label{simulation}
We first evaluated our model by calculating response time densities for two sets of parameters. 
These parameters were chosen to test the model in both a medium and a heavy load scenario. The expectation is to have higher accuracy in the case of medium load.
We inverted the Laplace transforms numerically using the Fixed Talbot inversion method to obtain the probability density functions \cite{AbaWhi95}.

\begin{figure}[ht]
    \centering
    \includegraphics[scale=0.8]{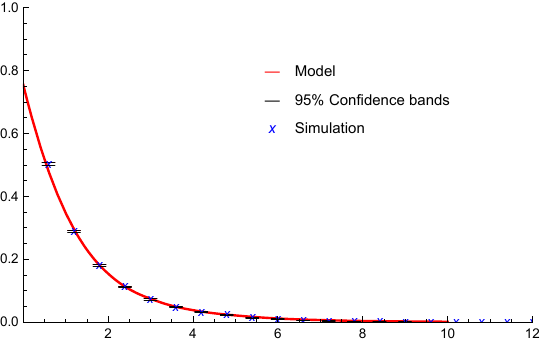}
    \caption{Response time density $w(t)$ (vertical axis) against $t$ (horizontal axis) for two JSQ-PS servers with parameters $\lambda=1, \mu_1=0.9, \mu_2=1.1, a_1=0, a_2=1$: comparison with simulation.}
    \label{jsq-ps-sim-vs-model-med}
\end{figure}

For the medium load scenario, the parameters used are $\lambda=1, \mu_1=0.9, \mu_2=1.1, a_1=0, a_2=1$, giving $50\%$ utilization. Notice that in this case all arrivals are directed to the faster server when the queue lengths are equal; this is discussed further in Subsection \ref{routing}.  We used $64$ points around the circle for both coordinates when solving for $E$ and $F$.
The density obtained was then plotted against corresponding simulation results. For the latter, we used regenerative simulation with 500,000 regeneration cycles; see \cite{Glynn2006}. Following \cite{Chien1995}, these sets of cycles were partitioned into batches, the sizes of which are chosen equal to the square root of the number of cycles, or nearest integer; here 707, giving a total of 707 batches as well.
Figure \ref{jsq-ps-sim-vs-model-med} shows that the model's accuracy is remarkable in this case. 
More details are given regarding the particular choice of routing probabilities $a_1, a_2$ in Subsection \ref{routing}.

\begin{figure}[ht]
    \centering
    \includegraphics[scale=0.2]{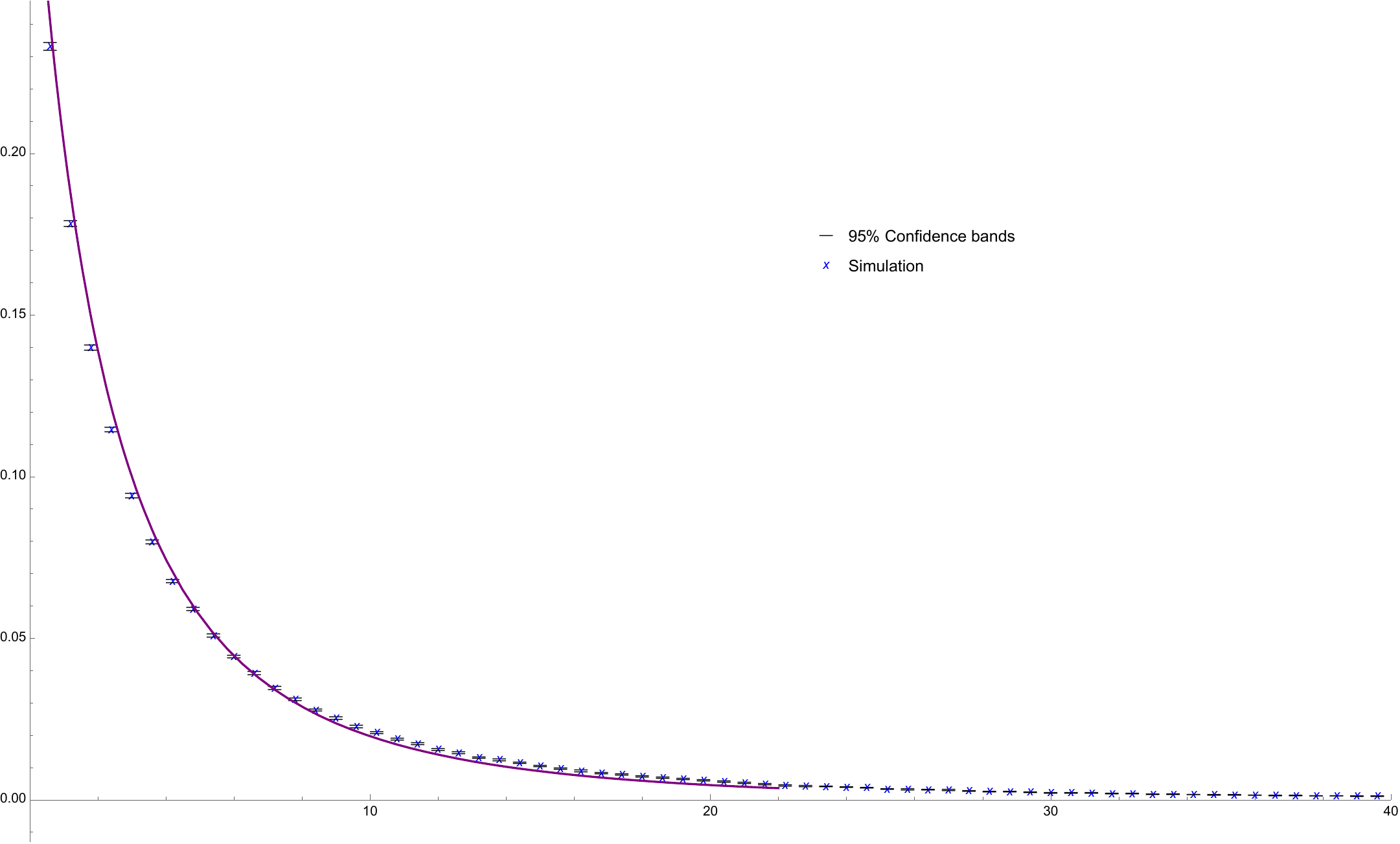}
    \caption{Response time density $w(t)$ (vertical axis) against $t$ (horizontal axis) for two JSQ-PS servers with parameters $\lambda=1, \mu_1=0.4, \mu_2=0.8, a_1=a_2=0.5$: comparison with simulation.}
    \label{jsq-ps-sim-vs-model-high}
\end{figure}

For the heavy traffic scenario, the parameters used are $\lambda=1, \mu_1=0.4, \mu_2=0.8, a_1=a_2=0.5$, so either queue would be saturated on its own since $\lambda>\mu_2>\mu_1$; the scheduling strategy is therefore more critical.
As before, we compared the results with the corresponding simulation. The densities are shown in Figure \ref{jsq-ps-sim-vs-model-high} and still show good agreement. Note, that the numbers of points used to calculate $E$ and $F$ were increased from $64$ to $120$ to achieve sufficient accuracy.

Finally, we used Lemma \ref{lem-moms} to compute response time moments up to the fourth by differentiation of $W^*(s)$ at $s=0$. This is much more efficient than calculating the response time densities as it only requires the model to run for a single $s$-value, namely $s=0$.  This is in contrast to estimating the density, where, at each time-point $t$, we need to calculate $W^*(s)$ for at least 16 $s$-values in order to be able to invert the LST with reasonable accuracy. 
Table \ref{moms-JSQPS} compares the first four moments of response time to those obtained from the corresponding regenerative simulation in the case of heavy traffic.
The model-results show very good accuracy for higher moments, well within the $95\%$ confidence intervals, and the first moment just outside.
Moments for the medium load scenario along with detailed discussion on optimal routing probabilities are given in Subsection \ref{routing}.

\subsection{Static scheduling models}\label{compare-to-static}
It is interesting to compare this JSQ model with other, static scheduling policies for the same two servers with total arrival rate $\lambda$. 
\begin{table}[t]
\caption{Response time moments for two JSQ-PS queues with parameters $\lambda=1$, $\mu_1=0.4$, $\mu_2=0.8$, $a_1=a_2=0.5$}
\begin{center}
\small
\begin{tabular}{cccc}
\toprule
Moment & JSQ/PS model & Simulation & 95\% CB \\
\midrule
1st & 6.15533 & 6.0914 & $\pm$ 0.022648 \\
2nd & 125.622 & 124.991 & $\pm$ 1.31098 \\
3rd & 5472.74 & 5547.67 & $\pm$ 128.633 \\
4th & 403965 & 422517 & $\pm$ 19382.9 \\
\bottomrule
\end{tabular}
\end{center}
\label{moms-JSQPS}
\end{table} 
The obvious policy would be to split the arrival stream so that each queue gets arrivals at rate proportional to its service rate; i.e. arrival rates $p_1\lambda$ and $p_2\lambda$ to queue 1 and queue 2, respectively, where $p_1=\mu_1/(\mu_1+\mu_2)$ and $p_2=\mu_2/(\mu_1+\mu_2)$.  The two queues then have known response time densities \cite{Coffman70}.  Notice that not any choice of probabilities $p_1,p_2$ give equilibrium, which requires $p_1\lambda<\mu_1$ and $p_2\lambda<\mu_2$; i.e. $1-\mu_2/\lambda<p_1<\mu_1/\lambda$.  Hence in our higher load case we require $0.2<p_1<0.4$.  The load balanced choice above has $p_1=\mu_1/(\mu_1+\mu_2)=1/3,\; p_2=2/3$ and yields equal utilizations 0.8333 at the two queues.
The response time in the static scheduling model has first four moments (obtained by differentiating the known Laplace transform of the density function) 10, 385.714, 3,5127.6 and 5,862,020, very much higher than those achieved by JSQ scheduling.

\begin{table}[t]
\caption{Comparison of response time moments for JSQ/PS, JSQ/FCFS, static load balanced with PS and static load balanced with FCFS; parameters: $\lambda=1, \mu_1=0.4, \mu_2=0.8, a_1=a_2=0.5$.}
\begin{center}
\small
\begin{tabular}{ccccc}
\toprule
Moment & JSQ/PS & JSQ/FCFS & Static/PS & Static/FCFS\\
\midrule
1st & 6.15533 & 6.07415 & 10.0 & 10.0 \\
2nd & 125.622 & 79.3979 & 385.714 & 225.0 \\
3rd & 5472.74 & 1678.18 & 35127.6 & 8437.5 \\
4th & 403965 & 49942. & 5862020 & 455625 \\
\bottomrule
\end{tabular}
\end{center}
\label{staticcfs}
\end{table}

\subsection{Finding the optimal routing probabilities}
\label{routing}

In this section, our goal is to find the value of $a_1$---the probability of an arriving task joining queue 1 given that the queues have equal length---that produces the best performance, or in other words, one that corresponds to the lowest moments. In the case of two identical queues, that is when $\mu_1=\mu_2$, it is straightforward to see that the optimal routing is $a_1=a_2=0.5$, otherwise, one of the servers becomes slightly overloaded. However, when the service rates are different, $a_1=a_2=0.5$ is not expected to be optimal.
One obvious candidate to try is splitting arrivals in proportion to each queue's service rate, similar to the static scheduling model above, making $a_1 = \frac{\mu_1}{\mu_1 + \mu_2}$.

The set of parameters used for this experiment is $\lambda=2, \mu_1=1, \mu_2=3$, and we take several values of $a_1$ between $0$ and $1$. The results are displayed in Table \ref{param123}.
The rows of Table \ref{param123} correspond to different values of $a_1$ and the moments computed using simulation are in brackets right next to their model-based counterpart.
Before focusing on the optimal row of Table \ref{param123}, one observation to make is that the moments calculated by the model are extremely accurate.

\begin{table}[ht]
\caption{Moments as the probability, $a_1$, of joining queue 1 at equal queue lengths varies;  model parameters: $\lambda=2, \mu_1=1, \mu_2=3$.}
\begin{center}
\tiny
\begin{tabular}{ccccc}
\toprule
    {Prob. of joining} &
      \multicolumn{4}{c}{Moments} 
\\ {queue 1, $a_1$} &
      \multicolumn{4}{c}{(in the form: \textit{model [simulation]})}
      \\
\midrule
0.0 & 0.731 [0.728] & 1.638 [1.637] & 7.455 [7.466] &53.204 [52.871] \\
0.1 & 0.761 [0.757] & 1.787 [1.777] & 8.461 [8.439]] & 62.185 [62.735] \\
0.25 & 0.813 [0.798] & 2.038 [1.983] & 10.099 [9.817] & 76.474 [74.304] \\
0.5 & 0.865 [0.858] & 2.360 [2.319] & 12.604 [12.228] & 101.432 [96.921] \\
0.75 & 0.920 [0.911] & 2.694 [2.643] & 15.215 [14.746] & 127.729 [121.712] \\
1.0 & 0.969 [0.957] & 3.009 [2.931] & 17.794 [17.045] & 154.705 [145.464] \\

\bottomrule
\end{tabular}
\end{center}
\label{param123}
\end{table}
\vspace{-0mm}
The $a_1$ value corresponding to the lowest moments is $a_1=0$. This suggests, perhaps not surprisingly, that when the service rates differ widely---in the current example queue 2 serves customers three times as fast as queue 1---the optimal strategy is to send everything to the faster server when there is a choice.

So next, we reran the experiments with parameters $\lambda=1, \mu_1=0.9, \mu_2=1.1$ to see if this is still the case when the service rates are much closer together; now queue 2 is only slightly more efficient than queue 1. Table \ref{param10911} shows the results. Again, the moments calculated by the model remain spot on.

\begin{table}[ht]
\caption{Moments as the probability, $a_1$, of joining queue 1 at equal queue lengths varies;  parameters: $\lambda=1, \mu_1=0.9, \mu_2=1.1$.}
\begin{center}
\tiny
\begin{tabular}{ccccc}
\toprule
    {Prob. of joining} &
      \multicolumn{4}{c}{Moments} 
\\ {queue 1, $a_1$} &
      \multicolumn{4}{c}{(in the form: \textit{model [simulation]})}
      \\
\midrule
0.0 & 1.399 [1.396] & 4.584 [4.574] & 25.778 [25.690] & 215.724 [214.135] \\
0.2 & 1.406 [1.407] & 4.640 [4.630] & 26.332 [26.079] & 222.872 [218.336] \\
0.4 & 1.434 [1.427] & 4.818 [4.773] & 27.765 [27.361] & 238.088 [232.156] \\
0.5 & 1.443 [1.431] & 4.883 [4.795] & 28.371 [27.468] & 245.269 [232.094] \\
0.6 & 1.452 [1.443] & 4.951 [4.902] & 29.019 [28.872] & 253.074 [257.151] \\
0.8 & 1.469 [1.457] & 5.094 [4.998] & 30.43 [29.414] & 270.534 [258.434] \\
1.0 & 1.486 [1.478] & 5.247 [5.199] & 32.021 [31.501] & 290.419 [280.881] \\

\bottomrule
\end{tabular}
\end{center}
\label{param10911}
\end{table}
\vspace{-5mm} The first row corresponding to $a_1=0$ remains the one with the lowest moments, suggesting that sending everything to the faster server is still the optimal strategy, even when service rates differ ever so slightly.

These findings prompt us to think about variations of JSQ scheduling itself. One can imagine a scheduling similar to JSQ where, given $\mu_1 < \mu_2$, arriving customers go to queue 1 if $i<j-1$ and to queue 2 if $i>j-1$, and routing probabilities are used to break ties when $i=j-1$.
More generally, we can shift the difference in the number of jobs by $k$, and route customers to the slower server when its queue is $k$ customers shorter than the faster server's queue. A natural question to ask is what is the optimal value of $k$ for a given pair of service rates.

\subsection{Accuracy and cost}
As per \cite{pgh2023}, we know the worst case errors for each of the matrices given in Table \ref{matrix-map}. The worst case errors of Table \ref{matrix-map-additional} are covered by the same propositions, since using Proposition \ref{prop:additional-mx}, the matrices can be expressed as linear combinations of matrices in Table \ref{matrix-map}. The operation with the largest asymptotic error corresponds to calculating derivatives with a worst case error of $\Theta(n^{-1})$ for each element of the matrix. 
Once the functional equations are transformed into a system of linear equations, the matrix on the left hand side is inverted. In our numerical calculations, the condition number of this matrix is low so that the error of the inverted matrix remains of the same order.
Finally, the inverted matrix is multiplied with a vector that is known exactly. Therefore, according to the central limit theorem, the error of the sum is of order $\sqrt{n}\Theta(n^{-1}) = \Theta(n^{-1/2})$. 
However, we found the error to be around 10-100 times smaller in practice; this is unsurprising because the above error estimates correspond to the worst case scenario, not the average. 

Regarding the cost of the algorithm, each of the matrices in equations \eqref{E-vec-rep-eq} and \eqref{F-vec-rep-eq} have size $n^2\times n^2$ and are calculated element-wise. Therefore, each requires storage and has computation time of order $\mathcal{O}(n^4)$. We used approximately $100$ points in each contour, so this is in the region of $10^8$.
However, these matrices need only be calculated once for any given $n$ and then can be stored and reused.

To get the final equations of form $\mat A(s)\vec x = \vec b$, a linear combination of the individual matrices in Table \ref{matrix-map} needs to be calculated, but these calculations are much faster than the calculations of the matrices in Table \ref{matrix-map}. However, for every time-point $t$ in the density function, we need approximately $16-32$ values of $s$, and corresponding matrices $\mat A(s)$, to be able to invert the Laplace transform numerically.  The product of 32 and the number of time-points required to approximate the density function well is significantly larger than $n$ in our calculations, so the overall cost exceeds $\Theta(n^5)$.
Next, the linear equations need to be solved for each $s$-point, which requires the inversion of the matrices $\mat A(s)$ -- a cubic operation in the dimension of the input matrix, in our case giving $\mathcal{O}(n^6)$ cost. This is the dominating factor, but in practice we found the time needed to be of the same order as calculating the $\mat A(s)$ matrices. This may be partly because $\mat A(s)$ requires the calculation of several matrices that each take $\Theta(n^5)$ time.
The remaining operations, e.g., inverting the Laplace transforms to obtain the densities and the calculation of moments, are negligible in comparison.


\section{Conclusion}\label{conc}
JSQ-PS is a widely used scheduling policy but its response time density -- even in the simplest case of two queues -- has not been obtained up until now.
We have used a novel numerical approximation technique to obtain the response time moments and density function for a pair of JSQ queues with PS discipline. 
The steps are as follows. First, functional equations for the generating functions $E$ and $F$ of conditional response times are derived, where the conditioning is on whether the arriving tagged task joins the first or the second queue. These functional equations are then solved numerically by transforming them into linear algebra. The generating function $G$ of the initial equilibrium queue-length probabilities is available from~\cite{pgh2023}.
Once the generating functions $G,E,F$ are approximated, together with their modifications (e.g. $E^{\sim}$), the evaluation of a complex integral is required to express the LST of the unconditional response time distribution, from which the density itself is obtained by numerical inversion. Moments are obtained very efficiently in a straightforward manner by recursively calculating derivatives of $W^*(s)$ at $s=0$. 
In section \ref{evaluation} we showed that our numerical method calculates the response time density and corresponding moments very accurately in both heavy and medium load.


Future work will explore transforming the functional equations by using $E'(x,y,s) = E(x,y,s)(1-x)(1-y)$ and similarly for $F$. It is expected that this transformation will increase accuracy as the right hand sides of the current functional equations have a pole at $(1,1)$ when $s=0$, requiring many more discretization points to achieve a high degree of accuracy when evaluating near the pole for contours with radii ($r_1, r_2$) not far from 1.
Furthermore, we also plan to explore variations on JSQ scheduling prompted by the experiments in Subsection \ref{routing} corresponding to finding the optimal values of $a_1$ and the offset parameter $k$. These variations can be thought of as shifted versions of the original JSQ scheduling policy, where the faster server automatically receives new arrivals until its queue is $k$ customers longer than the slower server's queue.

\bibliographystyle{IEEEtran}
\bibliography{references}

\newpage
\onecolumn

\section{Appendices}\label{Appendix}

\subsection{Mathematica code for two dimensional interpolation}
\label{2varinterp}
\label{}
\begin{figure}[ht]
    \centering
    \includegraphics[scale=0.6]{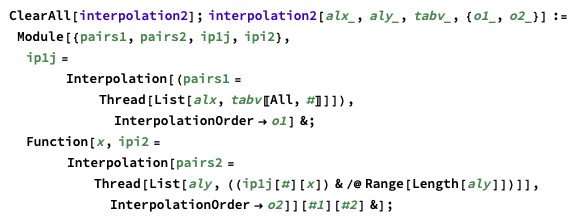}
    \caption{Mathematica code for 2-dimensional linear interpolation.}
    \label{fig:2dinterp}
\end{figure}

\subsection{Proof of Proposition \ref{prop:EandF}}\label{proofOfEandF}
\begin{proof}
Suppose that at time $t$, the tagged customer is in queue 1 with $i$ other customers and there are $j$ customer at queue 2.  Then, in the infinitesimal interval $(t,t+h]$ there are a number of possible events, one of which occurs with non-zero probability to first order in $h$:
\begin{enumerate}
    \item An arrival to queue 1, if $i+1<j$, with probability $\lambda h$;
    \item An arrival to queue 2, if $i+1>j$, with probability $\lambda h$;
    \item An arrival to queue 1, if $i+1=j$, with probability $a_1\lambda h$;
    \item An arrival to queue 2, if $i+1=j$, with probability $a_2\lambda h$;
    \item A departure from queue 2, if $j>0$, with probability $\mu_2 h$;
    \item A departure from queue 1 of a customer other than the tagged customer, if $i>0$, with probability $\frac{i}{i+1}\mu_1 h$;
    \item Completion of the tagged customer's sojourn with probability $\frac{1}{i+1}\mu_1 h$;
    \item No event occurs with probability $1-(\lambda+\mu_1+\mu_2I_{j>0})h$
\end{enumerate}
We therefore have the following equation to first order in $h$ for $i,j \geq 0$, with terms ordered according to the above list of events:
\begin{align*}
U_{i,j}(t+h) &= \mbox{} h\lambda I_{i+1<j} U_{i+1,j}(t) + h\lambda I_{i+1>j} U_{i,j+1}(t) + h\lambda a_1 I_{i+1=j} U_{i+1,j}(t) + h\lambda a_2 I_{i+1=j} U_{i,j+1}(t)\\
&+ h\mu_2 I_{j>0} U_{i,j-1}(t) + h\mu_1 I_{i>0} \frac i {i+1}U_{i-1,j}(t) + \frac{h \mu_1}{i+1} + (1-h(\lambda+\mu_1+\mu_2I_{j>0}))U_{i,j}(t)
\end{align*}
where $I$ is the indicator function, i.e. $I_B$ is 1 if $B$ is true and 0 otherwise. Rearranging, dividing by $h$ and taking the limit $h\rightarrow 0$ then yields
\begin{align*}
\frac{dU_{i,j}}{dt} = \mbox{} &\lambda I_{i+1<j} U_{i+1,j}(t) + \lambda I_{i+1>j} U_{i,j+1}(t) + \lambda a_1 I_{i+1=j} U_{i+1,j}(t) + \lambda a_2 I_{i+1=j} U_{i,j+1}(t)\\
&\mbox{} + \mu_2 I_{j>0} U_{i,j-1}(t) + \mu_1 I_{i>0} \frac i {i+1}U_{i-1,j}(t) + \frac{\mu_1}{i+1} - (\lambda+\mu_1+\mu_2I_{j>0})U_{i,j}(t)
\end{align*}
Taking the LST and multiplying throughout by $(i+1)$ we get
\begin{align*}
(s&+\lambda+\mu_1+\mu_2)(i+1)U^*_{i,j}(s)-\mu_2I_{j=0}(i+1)U^*_{i,0}(s)  =
\lambda I_{i+1<j} (i+1)U^*_{i+1,j}(s) + \lambda I_{i+1>j} (i+1)U^*_{i,j+1}(s)\\
&+ \lambda a_1 I_{i+1=j} (i+1)U^*_{i+1,j}(s) + \lambda a_2 I_{i+1=j} (i+1)U^*_{i,j+1}(s)+ \mu_2 I_{j>0} (i+1)U^*_{i,j-1}(s) + \mu_1 I_{i>0} i U^*_{i-1,j}(s) + \mu_1
\end{align*}
where $U_{i,j}^*(s)$ is the LST of the distribution function $U_{i,j}(t)$.  Multiplying by $x^iy^j$ and summing over $0\leq i,j < \infty$, with appropriate changes to these summation variables, yields (dropping the argument $s$ for brevity):

\begin{align*}
(s+&\lambda+\mu_1+\mu_2)\left( x \sum_{i=0}^\infty\sum_{j=0}^\infty U^*_{i,j} i x^{i-1} y^j + \sum_{i=0}^\infty\sum_{j=0}^\infty U^*_{i,j} x^i y^j \right) -
\mbox{} \mu_2 \left( x \sum_{i=0}^\infty U^*_{i,0} i x^{i-1} + \sum_{i=0}^\infty U^*_{i,0}x^i \right) = \\ 
&\lambda \sum_{i=1}^\infty\sum_{j=i+1}^\infty U^*_{i,j} i x^{i-1} y^j + \lambda \left( x \sum_{i=0}^\infty\sum_{j=0}^i U^*_{i,j+1} i x^{i-1} y^j + \sum_{i=0}^\infty\sum_{j=0}^i U^*_{i,j+1}x^i y^j \right) + \lambda a_1 \sum_{i=1}^\infty U^*_{i,i} i x^{i-1} y^{i} +\\
& \lambda a_2 y \left( x \sum_{i=0}^\infty U^*_{i,i+2} i x^{i-1} y^{i} + \sum_{i=0}^\infty U^*_{i,i+2}x^i y^{i} \right) + \mu_2 y \left(x\sum_{i=0}^\infty\sum_{j=0}^\infty U^*_{i,j} i x^{i-1} y^j + \sum_{i=0}^\infty\sum_{j=0}^\infty U^*_{i,j}x^i y^j\right) +\\
&\mu_1 x \left(x \sum_{i=0}^\infty\sum_{j=0}^\infty U^*_{i,j}i x^{i-1} y^j + \sum_{i=0}^\infty\sum_{j=0}^\infty U^*_{i,j} x^i y^j \right) +
\mu_1 \sum_{i=0}^\infty\sum_{j=0}^\infty x^i y^j
\end{align*}
Plugging in the definitions for $E(x,y),E^\sim(x,y),\beta_E^\sim(x,y)$ and $\delta_E^\sim(x,y)$ and moving everything but the last term to the left hand side, we get equation \ref{eqforE} as required.

The proof of equation \ref{eqforF} follows the exact same steps; it is omitted for brevity. 
\end{proof}

\subsection{Proof of Proposition \ref{wstarprop}} \label{proofwstar}
\begin{proof}
\begin{align*}
    W^*(s) &= \sum_{i=0}^\infty \sum_{j=0}^\infty \left(I_{i<j}U^*_{i,j}(s) + I_{i>j}V^*_{i,j}(s) + I_{i=j}a_1U^*_{i,j}(s) + I_{i=j}a_2V^*_{i,j}(s) \right)\pi_{i,j}\\
    &= \sum_{i<j} U^*_{i,j}(s)\pi_{i,j} + \sum_{i>j} V^*_{i,j}(s)\pi_{i,j} + a_1\sum_{i=j}^\infty U^*_{i,j}(s)\pi_{i,j}
    + a_2\sum_{i=j}^\infty V^*_{i,j}(s)\pi_{i,j}
\end{align*}
Omitting $s$ for brevity and using notation "$\sim$" as before for a binary operator, we focus on the $U^*_{i,j}$ terms first.
In order to be able to express $\sum_{i\sim j} U^*_{i,j}\pi_{i,j}$, we need to extract coefficients that satisfy $i=i'$ and $j=j'$ from the below sum
\begin{align*}
    E^\sim(x,y)G(z,q) = \sum_{i\sim j} \sum_{i'= 0}^\infty \sum_{j'=0}^\infty U^*_{i,j}\pi_{i',j'}x^i y^j z^{i'} q^{j'}
\end{align*}
This can be achieved by applying Lemma 1 in \cite{pghjb2021}, that is evaluating a certain  complex integral on a circle around the origin. Lemma 1 in \cite{pghjb2021} is stated in a more generic setting, here we provide an alternative, shorter argument for our specific case.
Given $r_1 \leq 1, r_2 = 1/r_1 \leq \frac{\mu_1+\mu_2}{\lambda}$, we have

\begin{align*}
   \frac{1}{(2\pi)^2} &\int_0^{2\pi} \int_0^{2\pi} E^\sim(r_1e^{\dotlessi t_1},r_1e^{\dotlessi t_2})G(r_2e^{-\dotlessi t_1},r_2e^{-\dotlessi t_2}) dt_1 dt_2\\
   &= \frac{1}{(2\pi)^2} \int_0^{2\pi} \int_0^{2\pi}  \sum_{i\sim j} \sum_{i'= 0}^\infty \sum_{j'=0}^\infty U^*_{i,j}\pi_{i',j'}r_1^i r_2^{i'}\left(e^{\dotlessi t_1}\right)^{i-i'} r_1^j r_2^{j'}\left(e^{\dotlessi t_2}\right)^{j-j'} dt_1 dt_2 \\
   &= \sum_{i\sim j} \sum_{i'= 0}^\infty \sum_{j'=0}^\infty U^*_{i,j}\pi_{i',j'}r_1^{i+j} r_2^{i'+j'}\left(\frac{1}{2\pi} \int_0^{2\pi} \left(e^{\dotlessi t_1}\right)^{i-i'} dt_1 \right) \left( \frac{1}{2\pi}\int_0^{2\pi} \left(e^{\dotlessi t_2}\right)^{j-j'} dt_2 \right)\\
    &= \sum_{i\sim j} U^*_{i,j}\pi_{i,j}(r_1 r_2)^{i+j} =  \sum_{i\sim j} U^*_{i,j}\pi_{i,j}
\end{align*}
where, and for the penultimate equality, we used the fact that 
\begin{equation*}
    \int_0^{2\pi} e^{\dotlessi t k} dt = 
    \begin{cases}
    0 \;\; \text{if} \;\; k \neq 0\\
    2\pi \;\; \text{if} \;\; k=0
    \end{cases}
\end{equation*}
The sums containing the $V^*_{i,j}$ terms are obtained similarly.
\end{proof}

\subsection{Proof of Lemma \ref{lem-moms}} 
\begin{proof}
First, as already noted, $\vec e_0 = \mat \Gamma_E \left(\frac{\mu_1}{(1-x_i)(1-y_j)}\,|\,1\leq i,j\leq n\right)^T$.
Differentiating equation \eqref{eqforE} $k$ times wrt $s$ (denoted by the superscript $(k)$), noting that the variable $s$ appears in only one coefficient and that the right hand side is constant, yields, for $k\geq 1$,
\begin{align*}
&\big(\lambda+(1-x)\mu_1+(1-y)\mu_2\big) \big(E^{(k)}(x,y,s)+x \frac{\partial E^{(k)}}{\partial x}(x,y,s)\big)
- \mu_2 \big(E^{(k)}(x,0,s)+x \frac{\partial E^{(k)}}{\partial x}(x,0,s)\big) \\
&- \lambda \frac{\partial E^{(k)<}}{\partial x}(x,y,s) - \lambda \big(\beta_E^{(k)\geq}(x,y,s) + x \frac{\partial \beta_E^{(k)\geq}}{\partial x}(x,y,s)\big) - \lambda a_1 \frac{\partial E^{(k)=}}{\partial x}(x,y,s)
 - \lambda a_2 y \big(\delta_E^{(k)=}(x,y,s) \\
 &+ x \frac{\partial \delta_E^{(k)=}}{\partial x}(x,y,s)\big) 
 = \mbox{} - k \big(E^{(k-1)}(x,y,s)+x \frac{\partial E^{(k-1)}}{\partial x}(x,y,s)\big)
\end{align*}

The left hand side is the left hand side of \eqref{eqforE} with $E$ replaced by $E^{(k)}$. Hence, as per the preceding discussion, if we set $r_E(x_i,y_j,s)=- k \big(E^{(k-1)}(x_i,y_j)+x_i \frac{\partial E^{(k-1)}}{\partial x}(x_i,y_j)\big)$, we find that $\vec e_k = \mat \Gamma_E \left(r_E(x_i,y_j,s)\,|\,1\leq i,j\leq n\right)^T$.  The result then follows for $\vec e_k$ using the operators in Table \ref{matrix-map}. The proof for $\vec f_k$ is similar.
\end{proof}

\end{document}